\documentclass[12pt]{article}
\usepackage{fullpage}
\usepackage{graphicx}
\newcommand{\be}{\begin{equation}}
\newcommand{\ee}{\end{equation}}

\renewcommand{\hat}{\widehat}
\renewcommand{\tilde}{\widetilde}
\def\x {{\bf x}}
\font\mybb=msbm10 at 11pt

\def\bb#1{\hbox{\mybb#1}}

\def\a {\eta}
\def\b {\zeta}

\newcommand{\news}{\setcounter{equation}{0}\quad}
\def\ben{\begin{equation}}
\def\een{\end{equation}}
\def\bea{\begin{eqnarray}}
\def\eea{\end{eqnarray}}
\begin{document}
\title{
\begin{flushright}\ \vskip -2cm {\normalsize{\em DCPT-10/09}}\end{flushright}
\vskip 2cm Skyrmions, instantons and holography}
\author{
Paul Sutcliffe\\[10pt]
{\em \normalsize Department of Mathematical Sciences,
Durham University, Durham DH1 3LE, U.K.}\\[10pt]
{\normalsize Email: 
 \quad p.m.sutcliffe@durham.ac.uk}
}
\date{May 2010}
\maketitle
\begin{abstract}
Some time ago, Atiyah and Manton observed that computing the
holonomy of Yang-Mills instantons yields good approximations to
static Skyrmion solutions of the Skyrme model. This paper provides
an extension and explanation of this result, by proving that 
instanton holonomies produce exact solutions of a BPS Skyrme model,
 in which the Skyrme field is coupled to a tower of vector mesons. 
Neglecting any (or indeed all) of the vector mesons breaks the
scale invariance and removes the BPS property of the Skyrmions.
However, it is shown that a truncation of the BPS Skyrme theory,
in which only the first vector meson is included, already
moves the Skyrme model significantly closer to the BPS system. 
A theory that is close to a BPS system is required to reproduce
the experimental data on binding energies of nuclei.    
A zero-mode quantization of the Skyrmion is performed in the truncated BPS
theory and the results are compared to the physical properties 
of the nucleon.
The approach is an analogue in five-dimensional Minkowski spacetime of 
a recent holographic construction of a Skyrme model 
by Sakai and Sugimoto,
based on a string theory derivation of a Yang-Mills-Chern-Simons theory 
in a curved five-dimensional spacetime.
\end{abstract}

\newpage
\section{Introduction}\news
The Skyrme model \cite{Sk} is a nonlinear theory of pions with
topological soliton solutions, called Skyrmions,
 that are identified as baryons. In this paper, three outstanding Skyrmion 
issues are drawn together and progress is made by 
 introducing an extended Skyrme model,
obtained from five-dimensional Yang-Mills theory. The three outstanding issues
that the new extended theory addresses are as follows.

Firstly, the Skyrme model is not a BPS (Bogomolny-Prasad-Sommerfield) theory,
in the sense that the soliton solutions do not attain the topological
lower bound on the energy. In fact, the single Skyrmion exceeds the
topological energy bound by $23\%$ in the case of massless pions.
 This energy excess allows the possibility of a significant classical 
binding energy for higher charge Skyrmions, and indeed this is the
case \cite{MSbook}:
 for example, the energy of the baryon number two Skyrmion exceeds
the topological bound by $18\%,$ which is already $5\%$ lower than
the single Skyrmion. Such binding energies are much
greater than those observed experimentally in nuclei, where binding
energies are typically less than $1\%.$ 
A BPS Skyrme model would therefore appear to be a better starting point
for obtaining more realistic binding energies, since a small perturbation
away from a BPS theory is likely to produce the required small 
binding energies. 
Motivated by this application a BPS Skyrme model is introduced, in which
the usual Skyrme model is extended by the inclusion of an infinite
tower of vector mesons. Furthermore, neglecting some of the vector
mesons provides a natural way to perturb away from the BPS system.

Secondly, Atiyah and Manton \cite{AM} have shown that computing the
holonomy of Yang-Mills instantons yields good approximations to
static Skyrmion solutions of the Skyrme model. It is not obvious
why this approximation turns out to be so successful, or if instanton
holonomies give exact solutions of some modified Skyrme model. 
This paper provides an explanation of the Atiyah-Manton procedure,
by proving that the holonomy of Yang-Mills instantons yields exact
solutions of the BPS Skyrme model. The BPS Skyrme model reverts to the 
usual Skyrme model if all the vector mesons are neglected, and this
explains the accuracy of the Atiyah-Manton approximation. Furthermore, 
it is shown that including only the first vector meson already 
significantly improves the accuracy of the instanton approximation 
of the Skyrme field and allows an excellent approximation to the 
vector meson field to be extracted from the instanton.

Thirdly, the usual Skyrme model includes pion degrees of freedom,
but neglects all the other mesons. There is a long history of attempts to
include other mesons, particularly the $\rho$ meson \cite{Ad,MZ}, but
there are difficulties because of the large number of coupling constants
that need to be determined:
although some progress has been made using
the ideas of hidden local symmetry and vector meson dominance 
\cite{BKUYY,BKY,HY}.
In the BPS Skyrme model, all parameters
 are uniquely determined once the energy and length
units are fixed. Truncating the BPS theory, for example by including only
the first vector meson, allows the usual Skyrme model to
be extended without the introduction of further unknown parameters.
The Skyrmion is studied in this truncated BPS theory, including its
zero-mode quantization, and the results are compared with the physical
properties of the nucleon. 

The techniques used in this paper, to derive the BPS Skyrme model and
its connection to instanton holonomies, are inspired by the work of
Sakai and Sugimoto \cite{SS}. Using a string theory construction and
holographic methods, they were able to derive a Skyrme model coupled
to an infinite tower of massive vector mesons from a Yang-Mills-Chern-Simons
theory in a curved five-dimensional spacetime. The Skyrme field of their
extended Skyrme model corresponds to the holonomy of the curved space
instanton, though unfortunately this instanton solution has not yet 
been determined, even numerically. The Skyrmion in the truncated version
of the extended Skyrme model, which includes only the pion and $\rho$ meson
degrees of freedom, has been investigated \cite{NSK}, though its quantization
has not. A collective coordinate quantization of the instanton has 
been performed \cite{HSSY,HRYYb}, but only
by approximating the true curved space Yang-Mills-Chern-Simons
 instanton by the flat space Yang-Mills instanton.

In some respects, 
the work in the present paper may be regarded as a five-dimensional
 Minkowski spacetime analogue of the five-dimensional curved 
spacetime theory of Sakai and Sugimoto \cite{SS}, with the advantage
that the instanton, and various other ingredients, can be
found explicitly. Of course, a disadvantage of this work is that
there is no AdS/CFT correspondence to justify the approach, even though 
several results are qualitatively similar to those of the Sakai-Sugimoto
theory, suggesting that there are some merits in considering this theory. 

\section{Skyrmions and instantons}\news
In the Skyrme model \cite{Sk} the pion
degrees of freedom are encoded into an $SU(2)$-valued Skyrme field
$U.$ In the massless pion approximation, the static energy of the Skyrme model is
\be
E_{\rm Sky}=\int \bigg(
-\frac{f_\pi^2}{4}\mbox{Tr}(R_iR_i)-\frac{1}{32e^2}\mbox{Tr}([R_i,R_j]^2)
\bigg)\,d^3x,
\label{energyfpi}
\ee
where $R_i=\partial_i U\,U^{-1}$ is the $su(2)-$valued current.
In the above, $e$ is the dimensionless Skyrme parameter and
$f_\pi$ may be interpreted as the pion decay constant. Note that there
are differing conventions, related by a factor of 2, for the pion decay
constant. In this paper the convention is chosen 
so that the physical value is $f_{\pi}=92.6\mbox{\,MeV},$ 
which agrees with the convention 
in \cite{SS} and related papers.

The parameters $f_\pi$ and $e,$ 
whose values are to be fixed by comparison with experimental data, 
merely set the energy and length units
and can be scaled away.
Explicitly, if energy units of $f_\pi/2e$ and length units of 
$1/ef_\pi$ are used, then in dimensionless Skyrme units the 
energy becomes
\be
E_{\rm Sky}=\int \bigg(
-\frac{1}{2}\mbox{Tr}(R_iR_i)-\frac{1}{16}\mbox{Tr}([R_i,R_j]^2)
\bigg)\,d^3x.
\label{energysu}
\ee
This dimensionless form is used in the remainder of this section. 

The Skyrme field is required to tend to a constant 
element of $SU(2)$ at spatial infinity 
(usually chosen to be the identity matrix) and this compactifies space to $S^3.$
A given Skyrme field therefore has an associated integer topological
charge $B\in\bb{Z}=\pi_3(SU(2))$ given explicitly by
\be
B=-\frac{1}{24\pi^2}\int\varepsilon_{ijk}\mbox{Tr}(R_iR_jR_k)\, d^3x.
\ee
It is this topological charge that is to be identified with baryon number \cite{Wi}. 
The Skyrmion of charge $B$ is the field $U$ that is the global minimum
of the energy (\ref{energysu}) for all fields in the given topological charge sector. 

The Faddeev-Bogomolny bound \cite{Fa} states that
\be
E_{\rm Sky}\ge 12\pi^2|B|,
\label{fb}
\ee
and it is easy to prove that this bound cannot be attained for 
non-zero $B.$

Recall that BPS solitons may be defined as solutions in which a 
topological energy bound is 
saturated and therefore, in this sense, Skyrmions are not BPS solitons.
Skyrmion solutions can only be obtained numerically and, 
as mentioned in the previous section, the energy of the $B=1$ Skyrmion
is $12\pi^2\times 1.23$ and the energy of the $B=2$ Skyrmion is
$24\pi^2\times 1.18.$ Numerical Skyrmion solutions have been obtained up to
reasonably large baryon numbers \cite{BS-full} and reveal that the
energies for larger values of $B$ are significantly closer to the bound than
for these low charge Skyrmions: for example, the $B=17$ Skyrmion has
an energy less than $17\times 12\pi^2\times 1.08.$ Computations based
on periodic Skyrme fields \cite{BS-lat} predict the limiting value 
$E_{\rm Sky}/B\rightarrow 12\pi^2\times 1.036,$ as $B\rightarrow\infty.$
The fact that the energy of the single Skyrmion is much further
from the bound than for larger values of $B$ implies binding
energies that are much greater than those found experimentally for nuclei,
which are typically less than $1\%.$ The small binding energies of nuclei
therefore motivate the search for a BPS Skyrme model, in which binding
energies would vanish, allowing the possibility that a small perturbation
of the BPS system might result in realistic nuclear binding energies. 

Although Skyrmions can only be obtained numerically, there are two 
analytic methods that produce Skyrme fields which are excellent approximations 
to the true Skyrmion solutions. One approach is the rational map approximation
\cite{HMS}, but the method of interest in this paper is that 
of Atiyah and Manton \cite{AM} in which a Skyrme field is generated
from the holonomy of a Yang-Mills instanton in $\bb{R}^4.$ This is
briefly reviewed below.

Let $A_I$ be the components of an $SU(2)$ Yang-Mills instanton in $\bb{R}^4,$
where uppercase latin indices run over all four space coordinates $I=1,2,3,4.$
The Skyrme field is defined to be the holonomy of this instanton 
computed along lines parallel to the $x_4$-axis. Explicitly,
\be
U(\x)=\pm{\cal P}\exp \int_{-\infty}^\infty A_4(\x,x_4)\,dx_4,
\label{amhol}
\ee
where ${\cal P}$ denotes path ordering and $\x=(x_1,x_2,x_3)$ are
 the Cartesian coordinates in the remaining 
$\bb{R}^3 \subset \bb{R}^4.$ As $A_4$ takes values in the Lie algebra
$su(2)$ its exponential is group-valued, so that
$U(\x):\bb{R}^3\mapsto SU(2)$, as required for a static Skyrme field.
The $\pm$ factor in (\ref{amhol}) is because the holonomy should really
be defined on a closed loop on $S^4$ and the sign may be required to
account for the transition function that connects $-\infty$ to $\infty,$
corresponding to the same point on $S^4.$ 

As shown by Atiyah and Manton \cite{AM}, the baryon number of this
Skyrme field is equal to the instanton number of the gauge field, 
that is, $B=N$ where 
 \be
N=-\frac{1}{16\pi^2}\int \mbox{Tr}(F_{IJ}\, ^\star F_{IJ})\, d^4x\,,
\label{ymtop}
\ee
and the dual field strength is defined by
$^\star F_{IJ}=\frac{1}{2}\varepsilon_{IJKL}F_{KL}.$

The Yang-Mills theory is conformally invariant and hence the instanton
field includes an arbitrary scale. This construction does not provide an exact 
solution of the Skyrme model for any instanton, but for each $N$ 
a suitable choice of instanton, including its scale, provides a remarkably good 
approximation to the static Skyrmion with baryon number $N.$
The energy of the approximate Skyrme field is typically around a percent 
higher than that of the numerical Skyrmion and correctly reproduces the 
symmetry of the Skyrmion for a range of highly symmetric cases
studied to date. For example, instantons have been constructed that 
correspond to the $B=1$ spherically symmetric and $B=2$ axially
symmetric Skyrmions \cite{AM}, tetrahedral and cubic Skyrmions
with $B=3$ and $B=4$ \cite{LM}, icosahedrally symmetric
Skyrmions with $B=7$ and $B=17$ \cite{SiSu,Su}, and the
triply periodic Skyrme crystal \cite{MS}. There is therefore significant
evidence to support the correspondence between Skyrmions and
instanton holonomies, though a deeper understanding of this 
connection and its remarkable accuracy is desirable.

Recently, the representation of a Skyrme field as an instanton
holonomy has reappeared in the context of five-dimensional theories
in compactified and/or curved spacetimes \cite{Hi,SoSt,SS,PW}. 
Although these approaches certainly have the flavour of the Atiyah-Manton
construction, none of them actually involve exact self-dual Yang-Mills
instantons in $\bb{R}^4.$ However, as the Sakai-Sugimoto 
construction \cite{SS} is a main motivation for the present paper, and
the techniques used here have some similarities to that work, it is perhaps
useful to give a brief overview of the relevant aspects of this model.
 
The Sakai-Sugimoto theory is based on a holographic approach to QCD in
the limit of a large number of colours, using the
AdS/CFT correspondence to map to a dual string theory consisting of 
probe D8-branes in a background of D4-branes. The action on the
probe D8-branes leads to a Yang-Mills-Chern-Simons theory in a 
five-dimensional curved spacetime. The spacetime involves a warped product
of (3+1)-dimensional Minkowski spacetime and an additional holographic 
direction. The static soliton in this theory is interpreted as the baryon,
 and has a fixed size determined by the ratio of the Chern-Simons
coefficient to the curvature associated with the holographic direction.
Unfortunately, the soliton in this theory has not been determined, 
even numerically. The expectation is that for a sufficiently small
Chern-Simons coupling the soliton size will be small enough that the
soliton can be approximated by a soliton of the flat space theory, which
is simply a self-dual Yang-Mills instanton in $\bb{R}^4,$ with a particular
small scale. In fact, all work to date on this theory has used the
flat space Yang-Mills instanton approximation, see for example \cite{HSSY,HRYYb}.
The dual theory in (3+1)-dimensional Minkowski spacetime is obtained
by performing an expansion of the five-dimensional theory in terms of
Kaluza-Klein modes of the holographic direction. These modes contain the
Skyrme field plus an infinite tower of massive vector mesons and the associated
theory is an extended Skyrme model, which reverts to the usual 
Skyrme model if the massive vector mesons are ignored. In summary, the
Sakai-Sugimoto theory provides a correspondence between 
Yang-Mills-Chern-Simons 
instantons on a curved four-manifold and an extended Skyrme model. 
One of the results of the present paper is to produce an analogous 
correspondence where the curved four-manifold is replaced by $\bb{R}^4$
and the Chern-Simons term is absent.
This allows a direct connection to be made to the Atiyah-Manton construction 
and also provides a natural extension of this method, together
with an understanding of the remarkable accuracy of this approach.   

It should be noted that a realization of the Atiyah-Manton 
construction has been proposed \cite{ENOT} in which instantons appear 
as domain wall Skyrmions in a five-dimensional Yang-Mills-Higgs theory.
This is certainly different from the approach discussed in the present paper
and does not involve vector mesons. It also implies that 
terms involving higher derivatives than the Skyrme term should appear,
which is not the case here, though it might be possible that some
connection could be made between the two approaches by integrating out the
vector mesons. 

\section{An abelian prototype}\news
As the details of the derivation for the full non-abelian gauge theory are
quite cumbersome, it is useful to first consider a similar approach
in a prototype abelian gauge theory, where the formulae are more manageable.

Consider an abelian gauge theory in five-dimensional Minkowski spacetime
with coordinates $t,x_I,$ where $I=1,2,3,4.$ For notational convenience
define $z=x_4$ and let lowercase latin indices run over the three
remaining spatial coordinates $i=1,2,3.$ 
The real-valued 
gauge potential has components $a_t,a_i,a_z.$
As most of this paper will be concerned with static fields with $a_t=0,$
attention may be restricted to the static Yang-Mills energy
\be 
E=\frac{1}{4} \int f_{IJ}f_{IJ}\ d^3x\, dz,
\label{aen}
\ee
where $f_{IJ}=\partial_I a_J-\partial_J a_I.$

As mentioned in the previous section, a crucial ingredient of the 
Sakai-Sugimoto construction \cite{SS} is the expansion of the gauge potential
 in terms of Kaluza-Klein modes in the holographic direction.
In flat Euclidean space a replacement needs to be found for the 
curved space Kaluza-Klein modes. Simply taking the zero curvature limit
is not suitable as the modes then degenerate to fourier modes, which
are not appropriate on the infinite line. The required modes must
form a complete orthonormal basis for square integrable functions on the 
real line with unit weight function (this is necessary to obtain
canonical kinetic terms for the vector mesons). This problem is familiar
to numerical analysts using spectral methods \cite{Bo} and the recognized
solution is provided by 
Hermite functions $\psi_n(z),$ where $n$ is a non-negative integer
and 
\be
\psi_n(z)=\frac{(-1)^n}{\sqrt{n!\,2^n\sqrt{\pi}}}e^{\frac{1}{2}z^2}
\frac{d^n}{dz^n}e^{-z^2}.
\label{hermite}
\ee
In a gauge in which $a_I\rightarrow 0$ as $|z|\rightarrow \infty,$ the
components of the gauge potential can be expanded in terms of
Hermite functions as
\be
a_z(\x,z)=\sum_{n=0}^\infty \alpha^n(\x) \psi_n(z), \quad\quad\quad
a_i(\x,z)=\sum_{n=0}^\infty \beta_i^n(\x) \psi_n(z).
\label{aseries}
\ee
Consider a gauge transformation $a_I\mapsto \tilde a_I=a_I-\partial_I h,$
for which $\tilde a_z=0.$ Clearly, this requires that $\partial_z h=a_z,$ and
hence $h$ is given by
\be
h(\x,z)=\int_{-\infty}^z a_{z}(\x,\xi)\, d\xi
=\sum_{n=0}^\infty \bigg(
\alpha^n(\x) \int_{-\infty}^z \psi_n(\xi)\, d\xi \bigg).
\label{agt}
\ee
Hermite functions satisfy
\be
\psi_n'(z)=\sqrt{\frac{n}{2}}\psi_{n-1}(z)
-\sqrt{\frac{n+1}{2}}\psi_{n+1}(z),
\label{dhermite}
\ee
where prime denotes differentiation with respect to $z.$
 This implies that 
\bea
\int_{-\infty}^z \psi_{2p+1}(\xi)\, d\xi
&=&\sum_{m=0}^{p} \gamma^m_{2p+1}\, \psi_{2m}(z),
\label{series1}\\
\int_{-\infty}^z \psi_{2p}(\xi)\, d\xi
&=&\gamma^+_{2p}\,\psi_+(z)+\sum_{m=0}^{p-1} \gamma^m_{2p}\, \psi_{2m+1}(z),
\label{series2}\eea
where $\gamma^+_{2p}$ and $\gamma^m_n$  are non-zero constants.

The additional function $\psi_+(z)$ has been introduced and is
defined by
\be
\psi_+(z)=
\frac{1}{\sqrt{2}\pi^\frac{1}{4}}\int_{-\infty}^z \psi_0(\xi)\, d\xi
=\frac{1}{2}+\frac{1}{2}\mbox{erf}(z/\sqrt{2})
\label{psiplus}
\ee
with $\mbox{erf}(z)$ the usual error function
\be
\mbox{erf}(z)=\frac{2}{\sqrt{\pi}}\int_0^z e^{-\xi^2}\, d\xi.
\label{erf}
\ee
The normalization of $\psi_+(z)$ has been chosen so that 
$\psi_+(-\infty)=0$ and $\psi_+(\infty)=1.$ 
 
The gauge transformation (\ref{agt}) can now be written in
terms of the basis functions $\psi_+(z),\psi_n(z)$ as
\be
h(\x,z)=u(\x)\,\psi_+(z)+\sum_{n=0}^\infty h^n(\x)\,\psi_n(z).
\label{agt2}
\ee  
As $\psi_n(\infty)=0$ and $\psi_+(\infty)=1,$ then
 $u(\x)$ is identified as the holonomy
\be
u(\x)=h(\x,\infty)=\int_{-\infty}^\infty a_{z}(\x,\xi)\, d\xi.
\ee
In the new gauge, where $\tilde a_z=0,$ then
\be
\tilde a_i=a_i-\partial_i h
=-\partial_i u(\x)\,\psi_+(z)+ 
\sum_{n=0}^\infty (\beta_i^n(\x)-\partial_i h^n(\x))\, \psi_n(z).
\ee
After defining the vector fields 
$v_i^n(\x)=\beta_i^n(\x)-\partial_i h^n(\x)$ this becomes
\be 
\tilde a_i=-\partial_i u(\x)\,\psi_+(z)+
\sum_{n=0}^\infty v_i^n(\x)\, \psi_n(z).
\label{aai}
\ee
In this gauge the holonomy appears in the boundary
condition $\tilde a_i\rightarrow -\partial_i u$ as $z\rightarrow\infty.$

Using (\ref{aai}) and (\ref{dhermite}) 
the components of the field strength are 
\bea
\tilde f_{zi}&=&-\partial_i u\,\psi_+'(z)+
\sum_{n=0}^\infty v_i^n\, \psi_n'(z) \\
&=&
\bigg(-\frac{1}{\pi^\frac{1}{4}}\partial_i u+v_i^1\bigg)
\frac{\psi_0(z)}{\sqrt{2}}
+\sum_{n=1}^\infty \bigg(v_i^{n+1}\,\sqrt{n+1}-v_i^{n-1}\sqrt{n}\bigg)
\frac{\psi_n(z)}{\sqrt{2}}\nonumber
\eea
and
\be \tilde f_{ij}=\sum_{n=0}^\infty (\partial_i v_j^n-\partial_j v_i^n)
\, \psi_n(z).
\ee
Using the orthonormality of the Hermite functions
\be
\int_{-\infty}^{\infty} \psi_m(z)\psi_n(z)\, dz=\delta_{mn}
\ee
to perform the integration over $z,$
the abelian Yang-Mills energy (\ref{aen}) becomes
\be
E=\int \bigg(
\frac{1}{4\sqrt{\pi}}(\partial_i u)^2
-\frac{1}{2\pi^\frac{1}{4}}v_i^1\partial_i u
+\sum_{n=0}^\infty\bigg\{
\frac{1}{4}(\partial_i v_j^n-\partial_j v_i^n)^2+\frac{1}{2}m_n^2(v_i^n)^2
-\frac{1}{2}q_nv_i^nv_i^{n+2}\bigg\}
\bigg)\ d^3x,
\label{aen2}
\ee
where the coefficients are $m_n^2={n+\frac{1}{2}}$ and $q_n=\sqrt{(n+1)(n+2)}.$

The second term in (\ref{aen2}) may seem a little strange, but it is 
simply the analogue in the prototype abelian theory of the familiar mixing
between the Skyrme field and the lightest axial vector meson that
arises in coupling the Skyrme model to vector mesons \cite{MZ}.  

This approach has produced a correspondence between abelian 
Yang-Mills theory in $\bb{R}^4$ and a field theory in $\bb{R}^3$ containing
an infinite tower of vector mesons, plus a scalar field related
to the holonomy of the gauge potential. 
Note that $m_n$ are not the meson masses because $q_n\ne 0,$ hence the
associated mass matrix is not diagonal. A truncated theory can be defined by 
including only the first ${\cal N}$ vector mesons and setting $v_i^n\equiv 0$ 
for all $n\ge{\cal N}.$ The mass matrix is then diagonalized by an
$SO({\cal N})$ rotation of the remaining vector mesons and the meson
masses determined from the eigenvalues of the ${\cal N}\times {\cal N}$ 
mass matrix. In the extreme case, ${\cal N}=1,$ where only the first vector 
meson remains, no rotation is required and the mass of this meson is 
obviously $m_0=1/\sqrt{2}.$ 

Perhaps it is worth making a comparison between the above approach and
the more common techniques of holographic QCD. In holographic QCD the
curvature of the extra dimension induces a discrete spectrum and fields
are then expanded in terms of the associated Kaluza-Klein modes. 
In the current situation the extra dimension is flat and therefore the
spectrum is continuous. A discrete spectrum must be identified 
in order to mimic the holographic construction. The traditional approach
to this problem in flat space is to compactify the extra dimension to
produce a discrete spectrum. The continuous spectrum is then recovered in
the limit of decompactification. However, for the application in the
present paper it is not appropriate to compactify the extra dimension, 
because a modification of space means that the connection to the instanton 
in $\bb{R}^4$ is then lost.
Furthermore, the identification of the holonomy with the
non-normalizable mode is no longer obvious in a compact extra dimension.
The above Hermite truncation selects a discrete spectrum without
the need to modify spacetime, and the continuous spectrum is recovered
in the limit ${\cal N}\rightarrow \infty.$ In this respect it is a 
crucial feature that the associated mass matrix is not diagonal. Increasing
${\cal N}$ not only adds an additional mode but also shifts the masses 
of all the 
previous modes, allowing the continuous spectrum to reappear 
as ${\cal N}\rightarrow \infty.$ 

In principle, the eigenfunctions of any Sturm-Lioville operator on
the line might be used as basis functions. However, a significant restriction
is imposed by the requirement that the basis functions are orthonormal with
respect to the unit weight function, which is needed to obtain the standard
kinetic terms for the vector mesons. Hermite functions are a canonical
choice and possess additional desirable features, such as 
the associated Sturm-Lioville operator involving only second and not 
first order derivatives. The precise combination of conditions that 
need to be imposed to uniquely arrive at the Hermite functions has
not been investigated.

In the next section a similar approach will be applied to $SU(2)$
Yang-Mills theory, where it is shown that the $SU(2)$-valued scalar field
that arises from the instanton holonomy is the Skyrme field, with
energy function precisely that of the Skyrme model.

\section{The Skyrme model from Yang-Mills theory}\news
Consider $SU(2)$ Yang-Mills gauge theory in $\bb{R}^4$ with the $su(2)$-valued
gauge potential $A_I,$ with $I=1,2,3,4,$ and again for convenience set $z=x_4.$ 
The Yang-Mills energy is given by 
\be
E=-\frac{1}{8}\int \mbox{Tr}(F_{IJ}F_{IJ})\,d^3x\,dz,
\label{yme}
\ee
where the factor of $\frac{1}{8}$ is due to the normalization of
the $su(2)$ generators as $-\mbox{Tr}(T_aT_b)=2\delta_{ab}.$

Starting with  a gauge in which $A_I\rightarrow 0$
as $|z|\rightarrow \infty,$ 
the gauge $A_z=0$ is obtained by applying the gauge transformation
\be
A_I\mapsto gA_I g^{-1}-\partial_I g\, g^{-1},
\ee
with
\be
g(\x,z)={\cal P}\exp \int_{-\infty}^z A_z(\x,\xi)\,d\xi.
\ee
The holonomy is 
\be
U(\x)=g(\x,\infty)={\cal P}\exp \int_{-\infty}^\infty A_z(\x,z)\,dz,
\ee
and in the new gauge the holonomy appears in the boundary condition 
for $A_i,$ since now
$A_i\rightarrow -\partial_iU\,U^{-1}$ as $z\rightarrow \infty.$ 

As in the abelian case (\ref{aai}), the gauge field can be expanded 
in terms of Hermite functions as
\be
A_i=-\partial_iU\,U^{-1}\,\psi_+(z)+
\sum_{n=0}^\infty W_i^n(\x)\, \psi_n(z).
\label{kk1}
\ee

The emergence of the Skyrme model can be seen by first neglecting
the vector fields $W_i^n.$
With this truncation the components of the field strength are
\be
F_{zi}=-\partial_iU\,U^{-1}\,\psi_+'=
-R_i\frac{\psi_0}{\sqrt{2}\pi^\frac{1}{4}},
\quad\quad\quad
F_{ij}=[R_i,R_j]\psi_+(\psi_+-1).
\ee
Substituting these expressions into the Yang-Mills energy (\ref{yme}),
and performing the integration over $z,$ yields the energy of the Skyrme model 
\be
E_{\rm S}=\int \bigg(
-\frac{c_1}{2}\mbox{Tr}(R_iR_i)-\frac{c_2}{16}\mbox{Tr}([R_i,R_j]^2)
\bigg)\,d^3x,
\label{skyen}
\ee
where
\be
c_1=\frac{1}{4\sqrt{\pi}}, \quad \quad 
c_2=\int_{-\infty}^\infty2\psi_+^2(\psi_+-1)^2\,dz=0.198.
\ee
This is the Skyrme model in dimensionless units, but it is 
not in Skyrme units because the constants $c_1$ and $c_2$ are not
unity.
In these units the Faddeev-Bogomolny energy bound (\ref{fb}) becomes
\be
E_{\rm S}\ge 12\pi^2\sqrt{c_1c_2}\,|B|= 2.005\,\pi^2\,|B|.
\ee
This bound should be compared with the energy bound that
derives from the full Yang-Mills theory, namely
\be
E\ge 2\pi^2\, |B|,
\label{ymbound}
\ee
which is attained by the instanton solutions. 
This shows that the two bounds are remarkably close, but that
the Faddeev-Bogomolny bound is stricter by $\frac{1}{4}\%.$
Of course, the Faddeev-Bogomolny bound only applies to the
usual Skyrme model, whereas the bound (\ref{ymbound})
is equally valid if some, or indeed all, of the vector mesons are included.

The Skyrme model is not scale invariant, in contrast to the Yang-Mills
theory, and hence the scale of the Skyrme model has emerged because
of the truncation that ignores the vector mesons. Perhaps it is useful to
think of this in terms of the theory flowing to a conformal theory
as all the vector mesons are included. 
The issue of Skyrmion and instanton scales also appears 
in the Sakai-Sugimoto derivation \cite{SS} of the Skyrme model. 
In that case the Yang-Mills theory in curved space with a Chern-Simons 
term is not scale invariant and the instanton has a fixed
small size, of the order of the string scale. The size of the Skyrmion
in the usual Skyrme model is not related to the size of the instanton, but as
more vector mesons are included the size of the Skyrmion in the extended
Skyrme model must tend to the small size of the instanton. 

The energy of a single Skyrmion exceeds the Faddeev-Bogomolny bound by about
$23\%$ and the energy of a Skyrme field generated from the holonomy
of a single instanton exceeds the Faddeev-Bogomolny bound by about $24\%,$
for an optimal choice of the instanton scale. When combined
with the above bounds, this reveals that, for the optimal instanton 
scale, the result of neglecting all the vector mesons is to raise the 
energy by less than $25\%.$ The vector meson terms in the expansion (\ref{kk1})
have trivial topology and therefore the holonomy term captures all the 
topological features of the instanton (and hence the Skyrmion), 
but the above results demonstrate that it also captures most of the 
energetic properties too.

Including the infinite tower of vector mesons produces a BPS Skyrme model,
since the model is simply equivalent to Yang-Mills theory with one extra dimension.
An infinite sequence of extended Skyrme models exist that interpolate
between the usual Skyrme model and the BPS Skyrme model, as the number 
of included vector mesons ranges from zero to infinity.
The remainder of this paper is devoted to a detailed analysis of the first
member of this sequence that extends the usual Skyrme model.

\section{Including the first vector meson}\news
This section considers the extension of the Skyrme model obtained by
including only the first vector meson, which physically
corresponds to coupling the pion field to the $\rho$ meson. 

The expansion (\ref{kk1}) is not convenient once the vector mesons
are included, because the fields do not have a definite parity.
It is first necessary to perform an additional gauge transformation
to obtain an expansion in terms of parity eigenstates. 

Given the holonomy $U,$ define the $SU(2)$-valued field $S$ to be
its positive square root, so that $S^2=U.$ 
After a gauge transformation by $g=S^{-1},$ the expansion
(\ref{kk1}) takes the form
\be
A_i=-\frac{1}{2}P_i\psi_\star+\frac{1}{2}Q_i+\sum_{n=0}^\infty V_i^n\psi_n
\label{expansion}
\ee
where $\psi_\star=\mbox{erf}(z/\sqrt{2})$ and
\be
P_i=S^{-1}\partial_i S+\partial_i S\,S^{-1},\quad\quad\quad
Q_i=S^{-1}\partial_i S-\partial_i S\,S^{-1}.
\ee
The vector meson $V_i^n$ is simply the previous vector meson $W_i^n$
in the new gauge.

Including only the first vector meson $V^0_i$ (and dropping the
superscript $0$ for notational convenience) gives
\be
F_{zi}=-P_i\frac{\psi_0}{\sqrt{2}\pi^\frac{1}{4}}
-V_i\frac{\psi_1}{\sqrt{2}}
\ee
and
\bea
F_{ij}&=&-[P_i,P_j]\frac{1}{4}(1-\psi_\star^2)
+(\partial_i V_j-\partial_j V_i)\psi_0
+[V_i,V_j]\psi_0^2
\nonumber\\
& &-([P_i,V_j]+[V_i,P_j])\frac{1}{2}\psi_\star\psi_0
+([Q_i,V_j]+[V_i,Q_j])\frac{1}{2}\psi_0.
\eea
Substituting these expressions into the Yang-Mills energy and
integrating over $z$ produces the energy
\be
E=E_{\rm S}+E_{\rm V}+E_{\rm I},
\label{ensvi}
\ee
where $E_{\rm S}$ is the earlier Skyrme energy (\ref{skyen}) and
$E_{\rm V}$ is the vector meson energy
\be
E_{\rm V}=\int -\mbox{Tr}\bigg\{
\frac{1}{8}(\partial_i V_j-\partial_j V_i)^2
+\frac{1}{4}m^2V_i^2
+c_3(\partial_i V_j-\partial_j V_i)[V_i,V_j]
+c_4[V_i,V_j]^2
\bigg\}\,d^3x,
\label{env}
\ee
with mass $m=\frac{1}{\sqrt{2}}$ and constants
\be
c_3=\int_{-\infty}^\infty
\frac{1}{4}\psi_0^3
\,dz=\frac{1}{2\sqrt{6}\pi^\frac{1}{4}},
\quad\quad
c_4=\int_{-\infty}^\infty
\frac{1}{8}\psi_0^4
\,dz=\frac{1}{8}\sqrt{\frac{1}{2\pi}}.
\ee 

In most phenomenological approaches to including the $\rho$ meson,
the energy $E_{\rm V}$ is taken to be that of a massive Yang-Mills
field. However, $E_{\rm V}$ only has this form if 
${2 c_3^2}/{c_4}$ is equal
to unity, whereas
${2 c_3^2}/{c_4}={2\sqrt{2}}/{3}=0.94,$
and hence there is a slight difference from a massive Yang-Mills theory.

The interaction energy $E_{\rm I}$ is
\bea
E_{\rm I}&=&\int -\mbox{Tr}\bigg\{
-c_5[P_i,P_j](\partial_i V_j-\partial_j V_i)
-c_6[P_i,P_j][V_i,V_j]
-c_5[P_i,P_j][Q_i,V_j]
\nonumber
\\
& &
+\frac{1}{4}[Q_i,V_j](\partial_i V_j-\partial_j V_i)
+c_3[Q_i,V_j][V_i,V_j]
+c_7([P_i,V_j]+[V_i,P_j])^2
\nonumber\\ & &
+\frac{1}{32}([Q_i,V_j]+[V_i,Q_j])^2
\bigg\}\,d^3x,
\eea
where the constants are
\bea
& &
c_5=\int_{-\infty}^\infty\frac{1}{16}(1-\psi_\star^2)\psi_0\,dz
=\frac{\pi^\frac{1}{4}}{12\sqrt{2}}
,\quad
c_6=\int_{-\infty}^\infty\frac{1}{16}(1-\psi_\star^2)\psi_0^2\,dz
=0.049
\nonumber\\
& &
c_7=\int_{-\infty}^\infty\frac{1}{32}\psi_\star^2\psi_0^2\,dz
=\frac{1}{32}-\frac{1}{2}c_6=0.007.
\eea 

The energy (\ref{ensvi}) corresponds to an extended Skyrme model in
which both the pion and $\rho$ meson degrees of freedom are included.
Models similar to this have been investigated in the past \cite{Ad,MZ,BKY,HY}
but because of the difficulty in determining the many possible interaction
coefficients some of these terms have not been included, and/or 
simple relations between the coefficients have been imposed, for example by
generating the theory using principles of hidden local symmetry \cite{BKUYY}.
An advatange of the current derivation is that all interaction coefficients
are uniquely fixed from the higher-dimensional Yang-Mills theory.

A similar truncation of the Sakai-Sugimoto theory produces an energy of
precisely the same form as that found here, but with different values for
the coefficients, and the classical single Skyrmion has been studied
numerically \cite{NSK}. 
The results show that the classical energy of the Skyrmion in that case
is reduced by about $10\%$ in comparison to the usual Skyrmion, but no
quantization of the Skyrmion has been performed.
In a later section it will be shown that the energy of a single Skyrmion
in the theory (\ref{ensvi}) is even lower than this, and moreover a
quantization of the Skyrmion reveals that it is vital to take into account
the quantum spin energy of the Skyrmion when attempting to match to
the physical properties of the nucleon. 

\section{The Skyrmion and its instanton approximation}\news
As the extended model (\ref{ensvi}) is obtained directly from Yang-Mills
theory then good approximations to the energy minimizing fields of
this model will be obtained by extracting the appropriate components
of the instanton, in terms of the basis expansion used for the truncation.
In this section this extraction is performed for the case of a single 
Skyrmion
through a detailed calculation of the decomposition of
the instanton in terms of the expansion (\ref{expansion}).
 
The single Skyrmion and the single instanton both have $SO(3)$ symmetry,
associated with spherical symmetry in $\bb{R}^3.$
It is therefore useful to introduce the following 
symmetric tensors
\be
X_{ia}=\delta_{ia}-\hat x_i\hat x_a, \quad
Y_{ia}=\hat x_i\hat x_a, \quad
Z_{ia}=\epsilon_{ija}\hat x_j,
\ee
where $\hat x_i=x_i/|\x|.$

The $N=1$ instanton located at the origin in $\bb{R}^4$ 
is given by $A_I=iA_{I a}\tau_a$ where
$\tau_a$ are the Pauli matrices and
\be
A_{ia}=\a(X_{ia}+Y_{ia})+\b Z_{ia}, \quad\quad\quad A_{za}=\b Y_{ia}, 
\ee
with the functions $\a$ and $\b$ defined to be
\be
\a(r,z)=\frac{z}{\lambda^2+r^2+z^2}, \quad\quad\quad
\b(r,z)=-\frac{r}{\lambda^2+r^2+z^2}.
\ee
Here $\lambda$ is the arbitrary scale of the instanton.
 
The gauge $A_z=0$ is obtained after the gauge transformation
\be
g(\x,z)=\mbox{exp}(iF(r,z)\hat x_a\tau_a),
\label{gt}
\ee
where 
\be
F(r,z)=\int_{-\infty}^z \b(r,\xi)\,d\xi=
-\frac{\pi r}{\sqrt{\lambda^2+r^2}}\bigg\{
\frac{1}{2}+\frac{1}{\pi}\mbox{tan}^{-1}
\bigg(\frac{z}{\sqrt{\lambda^2+r^2}}\bigg)\bigg\}.
\ee
The associated holonomy has the standard hedgehog form
\be
U=\mbox{exp}(if(r)\hat x_a\tau_a),
\label{hh}\ee
with profile function 
\be
f(r)=F(r,\infty)=-\frac{\pi r}{\sqrt{\lambda^2+r^2}}.
\label{hhp}
\ee
The Skyrme field must tend to a constant
element of $SU(2)$ as $r\rightarrow\infty$ and this is usually taken to be the identity
matrix, corresponding to the boundary condition $f(\infty)=0.$
In this section it is slightly more convenient to take the non-standard
choice that $U\rightarrow -1$ as $r\rightarrow\infty,$ with
the associated profile function boundary conditions $f(0)=0$ and $f(\infty)=-\pi,$
as satisfied by the profile function in (\ref{hhp}).
This is not an important change but does mean that some additional
factors of $\pi$ do not need to be introduced and carried throughout
the following calculation.

\begin{figure}[ht]\begin{center}
\includegraphics[width=10cm]{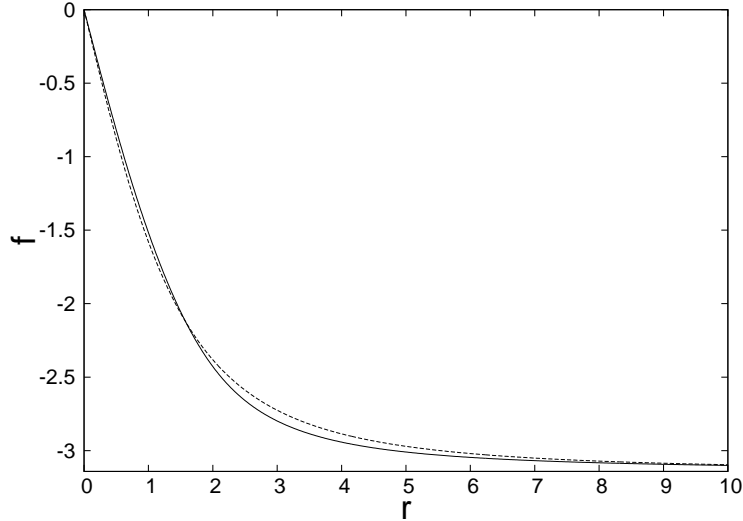}
\caption{The Skyrme field profile function minimizing the energy $E_{\rm S}.$
The solid curve is the numerical solution and the dashed curve is the 
instanton approximation.
}
\label{fig-pro1}\end{center}\end{figure}

First, consider restricting to the usual Skyrme energy $E_{\rm S},$
which will reproduce the results of Atiyah and Manton \cite{AM}.

For a Skyrme field of the hedgehog
form (\ref{hh}) the usual Skyrme energy (\ref{skyen}) reduces to
the expression
\be
E_{\rm S}=4\pi\int_0^\infty\bigg\{
c_1\bigg(f'^2+\frac{2\sin^2f}{r^2}\bigg)
+c_2\frac{\sin^2f}{r^2}\bigg(2f'^2+\frac{\sin^2f}{r^2}\bigg)\bigg\}\,r^2\,dr.
\ee
A numerical minimization of $E_{\rm S}$ 
yields an energy of  $E_{\rm S}=1.236\times 2\pi^2,$
with the associated profile function $f(r)$ displayed as the solid curve 
in  Figure~\ref{fig-pro1}. 

\begin{figure}[ht]\begin{center}
\includegraphics[width=10cm]{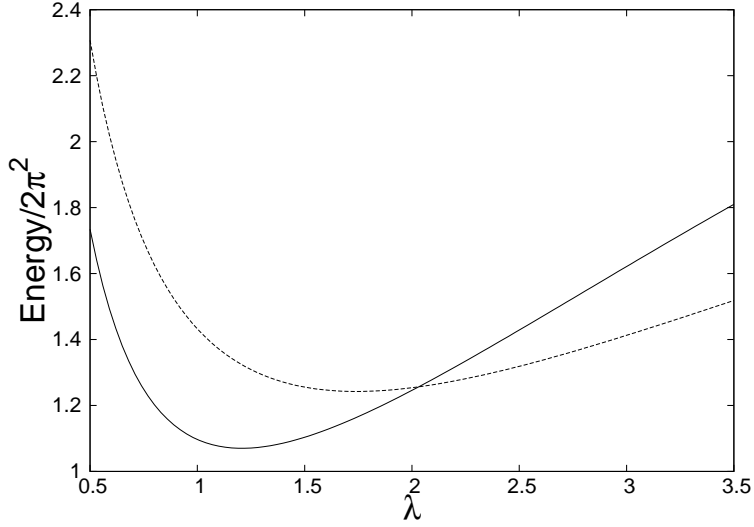}
\caption{The usual Skyrme energy $E_{\rm S}/2\pi^2$ (dashed curve) and 
the energy in the extended model $E/2\pi^2$ (solid curve)
 as a function of the instanton scale $\lambda.$
 }
\label{fig-en}\end{center}\end{figure}

Restricting to the instanton approximation
(\ref{hhp}),  the usual Skyrme energy $E_{\rm S}/2\pi^2$ 
is plotted as a function of the instanton scale $\lambda$ as
the dashed curve in Figure~\ref{fig-en}.
This computation 
reveals that within the instanton approximation $E_{\rm S}$  is minimized
by an instanton with scale $\lambda=1.72,$ which has an energy 
$E_{\rm S}=1.246\times 2\pi^2.$ 
For comparison, the instanton generated profile function with 
minimizing scale is displayed as the dashed curve in Figure~\ref{fig-pro1}. 
The instanton is found to generate a good approximation to the 
Skyrme field, with energy only $1\%$ above that of the numerical solution.

The next step is to extend
these calculations to determine the vector meson fields hidden inside the 
instanton. The energy minimizing Skyrme field and vector meson field are
then computed in the extended Skyrme model (\ref{ensvi}) and
compared with those from the instanton approximation.

Performing the gauge transformation (\ref{gt}) yields the gauge $A_z=0,$
and in this gauge the remaining components are
\be
A_{ia}=X_{ia}\bigg(
\a\cos 2F-(\b+\frac{1}{2r})\sin 2F\bigg)
+Y_{ia}\bigg(\a-\partial_r F\bigg)
+Z_{ia}\bigg(\a\sin 2F+(\b+\frac{1}{2r})\cos 2F-\frac{1}{2r}\bigg).
\ee
Now perform the additional gauge transformation with
\be
g=S^{-1}=\mbox{exp}(-\frac{i}{2}f\hat x_a\tau_a),
\ee
to obtain
\be
A_{ia}=X_{ia}\bigg(
\a\cos H-(\b+\frac{1}{2r})\sin H\bigg)
+Y_{ia}\bigg(\a-\frac{1}{2}\partial_r H\bigg)
+Z_{ia}\bigg(\a\sin H+(\b+\frac{1}{2r})\cos H-\frac{1}{2r}\bigg),
\label{paritya}
\ee
where 
\be
H=-\frac{2r}{\sqrt{\lambda^2+r^2}}\mbox{tan}^{-1}
\bigg(\frac{z}{\sqrt{\lambda^2+r^2}}\bigg).
\ee
The gauge potential (\ref{paritya}) now has the correct parity
properties to be compared with the expansion (\ref{expansion})
in terms of parity eigenstates. 
An immediate comparison yields that
\be
P_{ia}=X_{ia}\frac{1}{r}\sin f
+Y_{ia} \,f', \quad\quad\quad \mbox{and}\quad\quad\quad 
Q_{ia}=Z_{ia}\frac{1}{r}(\cos f-1),
\ee
where $P_i=iP_{ia}\tau_a$ etc. Then, by definition of the
terms in the expansion (\ref{expansion}), this gives that
\be
V_{ia}^n=k_1^nX_{ia}+k_2^n Y_{ia}+k_3^n Z_{ia},
\ee
where the profile functions $k_i^n(r)$ are given by the integrals
\bea
k_1^n(r)&=&\int_{-\infty}^\infty\bigg\{
\a\cos H-(\b+\frac{1}{2r})\sin H+
\frac{1}{2r}\sin f\,\psi_\star
\bigg\}\,\psi_n\,dz\\
k_2^n(r)&=&\int_{-\infty}^\infty\bigg\{
\a-\frac{1}{2}\partial_r H+\frac{1}{2}f' \,\psi_\star\bigg\}
\psi_{n}\,dz\nonumber \\
k_3^n(r)&=&\int_{-\infty}^\infty\bigg\{
\a\sin H+(\b+\frac{1}{2r})\cos H-\frac{1}{2r}\cos f
\bigg\}\,\psi_n\,dz.\nonumber 
\eea

For the even vector mesons it is easy to show that
$k_1^{2n}(r)=k_2^{2n}(r)=0$
due to the symmetry $\psi_{2n}(-z)=\psi_{2n}(z)$, and 
for the odd vector mesons
$k_3^{2n+1}(r)=0$
due to the symmetry $\psi_{2n+1}(-z)=-\psi_{2n+1}(z).$
This is the correct parity associated with the fact 
that $V_i^n$ is a
vector meson for even $n$  and an axial vector meson for odd $n.$

For the first vector meson $n=0,$ and again dropping the superscript 
on $V^0_i,$ the above results reduce to
\be
V_{ia}=\rho(r)Z_{ia},
\label{rho}
\ee
where the profile function $\rho(r)$ is
\be
\rho(r)=\int_{-\infty}^\infty\bigg\{
\a\sin H+(\b+\frac{1}{2r})\cos H-\frac{1}{2r}\cos f
\bigg\}\,\psi_0\,dz.
\label{hhrho}
\ee
The profile function $\rho(r)$ satisfies the boundary conditions
$\rho(0)=\rho(\infty)=0.$

\begin{figure}\begin{center}
\includegraphics[width=10cm]{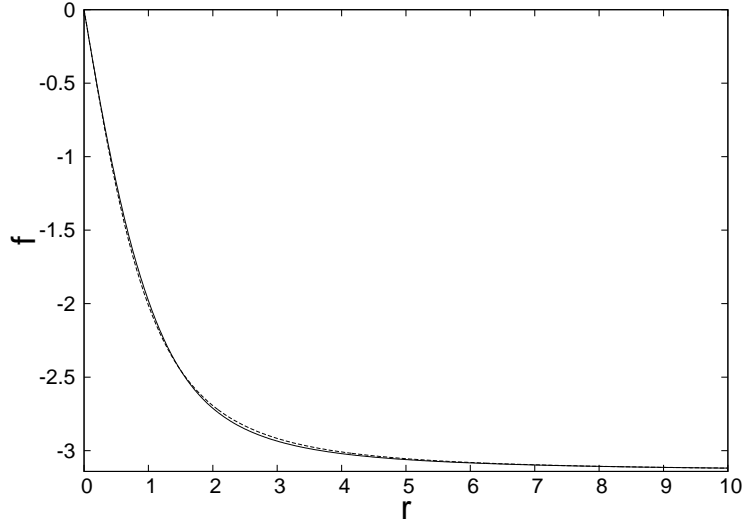}
\caption{The Skyrme field profile function minimizing the energy $E.$
The solid curve is the numerical solution and the dashed curve is the 
instanton approximation.
}
\label{fig-pro2}\end{center}\end{figure}

\begin{figure}\begin{center}
\includegraphics[width=10cm]{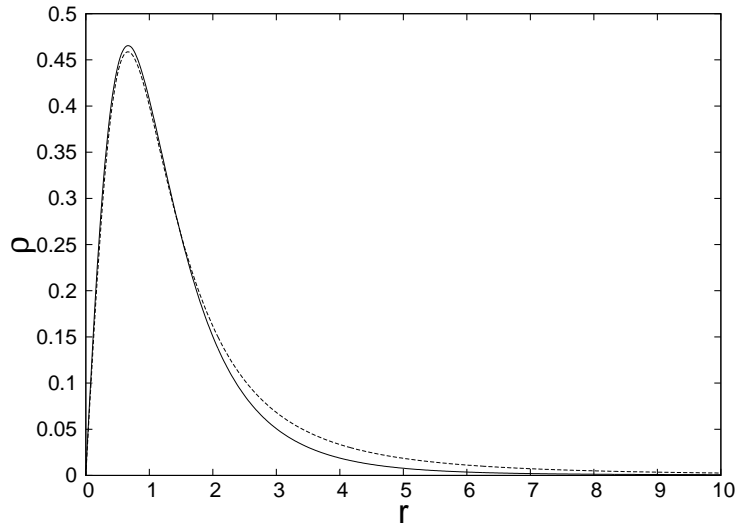}
\caption{The vector meson profile function $\rho$ 
minimizing the energy $E.$
The solid curve is the numerical solution and the dashed curve is the 
instanton approximation.
}
\label{fig-pro3}\end{center}\end{figure}

In terms of arbitrary profile functions $f(r)$ and $\rho(r)$ 
appearing in the spherical ansatz (\ref{hh}) and (\ref{rho}), the 
additional terms in the energy
$E=E_{\rm S}+ E_{\rm V}+ E_{\rm I}$
become
\be
E_{\rm V}=4\pi\int_0^\infty\bigg\{
\rho'^2+\frac{3\rho^2}{r^2}+\frac{2\rho\rho'}{r}
+m^2\rho^2
+c_3 16\frac{\rho^3}{r}+c_4 16\rho^4
\bigg\}\,r^2\,dr,
\ee
and
\bea
E_{\rm I}&=&4\pi\int_0^\infty\bigg\{
-16c_5\frac{\sin f}{r}\bigg(f'(\rho'+\frac{\rho}{r})
+\frac{\rho \sin f\cos f}{r^2}
\bigg)
+4\frac{\rho^2}{r^2}(\cos f-1)
\\& & 
-16c_6\frac{\rho^2\sin^2 f}{r^2}
+16c_3\frac{\rho^3}{r}(\cos f-1)
+32c_7 f'^2\rho^2
+2\frac{\rho^2}{r^2}(\cos f-1)^2
\bigg\}\,r^2\,dr.
\nonumber
\eea
A numerical minimization of the extended energy $E$ yields
the value $E=1.060\times 2\pi^2.$ This shows that the
energy of a Skyrmion in this theory is significantly closer to the
topological lower energy bound (\ref{ymbound}) than in the usual 
Skyrme model. This result reveals that the truncation of the BPS Skyrme
theory, in which only the first vector meson is included,
already moves the usual Skyrme model significantly closer to the BPS theory. 
The numerically determined Skyrme profile function $f(r)$ is displayed as the
solid curve in Figure~\ref{fig-pro2} and the vector meson profile
function $\rho(r)$ as the solid curve in Figure~\ref{fig-pro3}.

Applying the instanton approximation, with profile functions
(\ref{hhp}) and (\ref{hhrho}),  the energy $E/2\pi^2$ is plotted,
 as a function of the instanton scale $\lambda,$ 
as the solid curve in Figure~\ref{fig-en}.
The mimimizing instanton 
scale is $\lambda=1.20$ at which the energy is $E=1.071\times 2\pi^2.$ 
This demonstrates that the instanton
generated fields provide an excellent approximation to both the Skyrme field
and the vector meson field, with an energy only $1\%$ above that of the
numerical solution. For comparison to the numerical solutions, the
minimizing instanton profiles are displayed as the dashed curves
in Figure~\ref{fig-pro2} and Figure~\ref{fig-pro3}. 
Comparing Figure~\ref{fig-pro1} with Figure~\ref{fig-pro2}
confirms that, as expected, the instanton approximation to the Skyrme field profile
function is much closer to the numerical solution in the extended Skyrme
theory than in the usual Skyrme model: in fact, it is difficult to 
distinguish the two curves in Figure~\ref{fig-pro2}.

\section{Quantization of the Skyrmion}
It is well-known that fixing the parameters of the Skyrme model
in the meson sector does not produce good results in the baryon
sector.
It is common practice to treat (at least some of) the meson constants
as free parameters which are then fixed by comparison to selected
baryon properties \cite{ANW,AN,Ad,Ko,MW,BMSW}.
It has been suggested \cite{BKS} that the meson parameters could be
interpreted as renormalized constants in the baryon sector, that result
from known quantum effects not addressed within a simple zero-mode
quantization. As demonstrated below, 
the same situation persists in the extended Skyrme model.

Given the dimensionless formulation of the theory (\ref{ensvi}),
the energy unit $\varepsilon$ and the length unit $l$ are related
to the pion decay constant $f_{\pi}$
and $\rho$ meson mass $m_\rho$
by
\be
\varepsilon=f_\pi^2\sqrt{2\pi}/m_\rho, \quad\quad\quad 
l=1/(\sqrt{2}m_\rho).
\label{units}
\ee
This can be seen by performing the scaling 
$E\mapsto \varepsilon E,$ $x_i\mapsto x_i/l,$ 
$V_i\mapsto V_i\sqrt{{l}/{\varepsilon}},$ after which the 
dimensionless Skyrme energy (\ref{skyen}) takes the standard form 
(\ref{energyfpi}) 
with Skyrme parameter 
\be
e=\frac{m_\rho}{f_\pi\sqrt{2c_2}\pi^\frac{1}{4}}.
\label{skyrmee}\ee
The vector meson energy (\ref{env}) becomes
\be
E_{\rm V}=\int -\mbox{Tr}\bigg\{
\frac{1}{8}(\partial_i V_j-\partial_j V_i)^2
+\frac{1}{4}m_\rho^2V_i^2
+\tilde c_3(\partial_i V_j-\partial_j V_i)[V_i,V_j]
+\tilde c_4[V_i,V_j]^2
\bigg\}\,d^3x,
\ee
where
\be
\tilde c_3
=\frac{m_\rho}{2\sqrt{6\pi}f_\pi},
 \quad\quad\quad
\tilde c_4
=\frac{m_\rho^2}{8\pi\sqrt{2}f_\pi^2}.
\ee
For completeness, the additional coefficients in the interaction
energy $E_{\rm I}$ remain
unchanged except for the replacement
\be
c_5\mapsto\tilde c_5
=\sqrt{\frac{\pi}{2}}\frac{f_\pi}{12m_\rho}.
\ee

The classical Skyrmion energy is $E=\varepsilon M,$
where $M=1.06\times 2\pi^2$ is the dimensionless static energy. 
Taking the physical values
$f_{\pi}=92.6\mbox{\,MeV}$ and 
$m_\rho=776\mbox{\,MeV},$ to set the units (\ref{units}),
gives a classical Skyrmion energy $E=580\mbox{\,MeV},$
which is far too low in comparison to the nucleon mass of 
$939\mbox{\,MeV}.$ 
There is also a quantum contribution to the
energy associated with the spin of the nucleon, but even if this
contribution raised the total energy to that of the nucleon mass this
would still not be an acceptable result, since physically
the spin contribution needs to provide only a small contribution to
the total energy.
The quantum spin energy is calculated
later in this section and reveals that taking the physical values for
the meson parameters
$f_\pi$ and $m_\rho$ yields a quantum spin contribution of $1039\mbox{\,MeV},$
which is almost twice that of the classical energy and gives a total energy
which is far too large. This shows that taking the physical values for the
meson parameters does not produce acceptable results for the baryon.
Confirmation of this is provided by calculating the size of the baryon,
as follows.

The physical value of the nucleon
isoscalar root mean square radius is
$\sqrt{<r^2>}=0.72\mbox{\,fm}.$ For the Skyrmion its dimensionless
form $R$ is calculated from the radial baryon density ${\cal B}$ as
\be
R^2=\int_0^\infty r^2 {\cal B}\, dr=
-\int_0^\infty r^2 \frac{2}{\pi}f'\sin f\, dr=0.82.
\ee
Inserting the length unit gives
\be
\sqrt{<r^2>}=\frac{R}{\sqrt{2}m_\rho}.
\label{size}
\ee
Taking the physical value for $m_\rho,$ and using
the fact that in natural units $\mbox{MeV}^{-1}=197\mbox{\,fm}$,
yields $\sqrt{<r^2>}=0.16\mbox{\,fm},$ which is far too small.
This is the origin of the excessive quantum spin energy mentioned above, 
when physical values are taken for the meson parameters. The baryon is
far too small and hence so is its moment of inertia, which occurs in
the denominator of the quantum spin energy.

From now on the common practice is adopted of treating
the meson parameters of the theory (in this case $f_\pi$ and $m_\rho$) 
as free parameters that are to be fixed by comparison to physical 
properties in the baryon sector. 

The choice made in this paper is to fix the energy and length units
by matching to the physical values of the 
nucleon mass and the isoscalar root mean square 
radius. Matching the latter, using equation (\ref{size}), yields
$m_\rho=176\mbox{\,MeV},$ which is therefore only around a quarter of
its physical value. To determine $f_\pi$ it is first necessary to
calculate the quantum spin contribution to the nucleon mass, which
is presented below.

So far in this paper the discussion has been restricted to static fields.
It is a simple matter to obtain the relevant Lagrangians from
the static energies presented earlier by applying the obvious 
relativistic generalization. From the dimensionless form (\ref{ensvi}) 
of the static energy of the extended
Skyrme model, the associated dimensionless kinetic energy is 
$T=T_{\rm S}+T_{\rm V}+T_{\rm I}$ 
where
\be T_{\rm S}=
\int -\mbox{Tr}\bigg\{
\frac{c_1}{2}R_0^2
+\frac{c_2}{8}[R_0,R_i]^2
\bigg\}\,d^3x,\ee

\bea T_{\rm V}&=&\int-\mbox{Tr}\bigg\{
\frac{1}{4}(\partial_0 V_i-\partial_i V_0)^2
+\frac{1}{4}m^2V_0^2 
+2c_3(\partial_0 V_i-\partial_i V_0)[V_0,V_i]
+2c_4[V_0,V_i]^2
\bigg\}\,d^3x,
\nonumber\\
\eea

\bea T_{\rm I}&=&\int -\mbox{Tr}\bigg\{
-2c_5[P_0,P_i](\partial_0 V_i-\partial_i V_0)
-2c_6[P_0,P_i][V_0,V_i]
-c_5[P_0,P_i]([Q_0,V_i]+[V_0,Q_i])
\nonumber
\\
& &
+\frac{1}{4}([Q_0,V_i]+[V_0,Q_i])(\partial_0 V_i-\partial_i V_0)
+c_3([Q_0,V_i]+[V_0,Q_i])[V_0,V_i]
\nonumber\\ & &
+2c_7([P_0,V_i]+[V_0,P_i])^2
+\frac{1}{16}([Q_0,V_i]+[V_0,Q_i])^2
\bigg\}\,d^3x.\eea

The zero-mode quantization involves the rigid rotor ansatz
\be
S=e^{\frac{1}{2}\Omega t} \tilde S e^{-\frac{1}{2}\Omega t}, \quad\quad
V_i=e^{\frac{1}{2}\Omega t} \tilde V_i e^{-\frac{1}{2}\Omega t}, \quad\quad
V_0=e^{\frac{1}{2}\Omega t} \tilde V_0 e^{-\frac{1}{2}\Omega t},
\ee
where $\Omega=i\Omega_a\tau_a$ is a constant element of $su(2)$ 
determining the rotation frequency and axis, and 
$\tilde S,$ $\tilde V_i$  are the earlier static fields
that minimize the classical static energy $E.$ 
For time-dependent fields $\tilde V_0$ can no longer be set to zero,
as this is not consistent with the Gauss law for this system, that is, the
field equation for $V_0.$ The Gauss law requires that $\tilde V_0$
has the same tensorial structure as $[\tilde V_i,[\tilde V_i,\Omega]].$
This determines the form of $\tilde V_0=i\tilde V_{0a}\tau_a$ to be \cite{Ad}
\be
\tilde V_{0a}=\chi_1\Omega_a+\chi_2\hat x_a\hat x_b\Omega_b,
\ee
where $\chi_1(r),\chi_2(r)$ are two additional radial profile functions with
boundary conditions $\chi_1'(0)=\chi_1(\infty)=\chi_2(0)=\chi_2(\infty)
=0.$
The kinetic energy takes the form $T=\frac{1}{2}I|\Omega |^2,$ where
$|\Omega|^2=-\frac{1}{2}\mbox{Tr}(\Omega^2)$ and 
$I=I_{\rm S}+I_{\rm V}+I_{\rm I}$ is the moment of inertia, which
after a tedious calculation is found to be 
\be
I_{\rm S}
=\frac{16\pi}{3}
\int_0^\infty\bigg\{c_1+c_2\bigg(f'^2+\frac{\sin^2f}{r^2}\bigg)\bigg\}
\sin^2f \,r^2\,dr,
\ee

\bea
I_{\rm V}
&=&\frac{4\pi}{3}
\int_0^\infty\bigg\{4\rho^2+4\frac{\chi_2^2}{r^2}+3\chi_1'^2
+2\chi_1'\chi_2'+\chi_2'^2
+m^2(3\chi_1^2+2\chi_1\chi_2+\chi_2^2)
\nonumber\\& &
+32c_3\rho(2\rho\chi_1+\rho\chi_2+\frac{\chi_2^2}{r})
+64c_4\rho^2(2\chi_1^2+2\chi_1\chi_2+\chi_2^2)
\bigg\}
 \,r^2\,dr,
\eea

\bea
& &I_{\rm I}
=\frac{4\pi}{3}
\int_0^\infty\bigg\{
-32c_5\sin f\bigg(\frac{\sin f}{r}(\rho-\frac{\chi_2}{r})+f'\chi_1'\bigg)
-64c_6\rho \frac{\sin^2f}{r}\chi_1
\nonumber\\& &
-8\sin^2\frac{f}{2}\bigg(
-8c_5\frac{\sin^2f}{r}(\rho+\frac{\chi_1}{r})
+\rho^2+2\rho\frac{\chi_1}{r}+\frac{\chi_2^2}{r^2}
+8c_3\rho(2\frac{\chi_1^2}{r}+2\frac{\chi_1\chi_2}{r}
+\frac{\chi_2^2}{r}+\chi_1\rho)
\bigg)
\nonumber\\& &
+64c_7\bigg(
f'^2\chi_1^2+\sin^2f\bigg(
(\rho-\frac{\chi_1}{r})^2+\frac{(\chi_1+\chi_2)^2}{r^2}\bigg)
\bigg)
\nonumber\\& &
+8\sin^4\frac{f}{2}\bigg(
(\rho+\frac{\chi_1}{r})^2+\frac{(\chi_1+\chi_2)^2}{r^2}\bigg)
\bigg\}
 \,r^2\,dr.
\eea

\begin{figure}\begin{center}
\includegraphics[width=10cm]{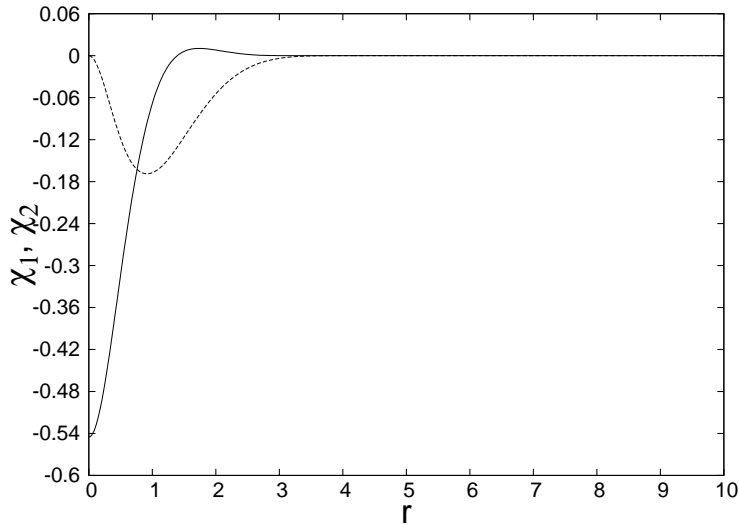}
\caption{The profile functions $\chi_1$ (solid curve) 
and $\chi_2$ (dashed curve).
}
\label{fig-chi}\end{center}\end{figure}

The functions $\chi_1,\chi_2$ are determined by the $V_0$ field equation
and this is equivalent to the minimization of $I_{\rm V}+I_{\rm I},$
given the profile functions $f$ and $\rho.$
The minimizing profile functions are presented in Figure~\ref{fig-chi}
and the associated moment of inertia is computed to be
$I=I_{\rm S}+(I_{\rm V}+I_{\rm I})=13.73+1.96=15.69$

In terms of the spin $J=I|\Omega|$ the dimensionless quantum spin energy is
\be
E_Q=\frac{J^2}{2I},
\ee
where $J^2=j(j+1)$ and the quantum spin number $j=\frac{1}{2}$ for the
nucleon. 
The moment of inertia $I$ has units of $\varepsilon l^2$ and the 
quantum spin energy has units which are the reciprocal of this.
The mass of the nucleon is therefore
\be
M_N=\varepsilon M + \frac{1}{\varepsilon l^2}E_Q
=\frac{f_\pi^2}{m_\rho}\sqrt{2\pi} M
+\frac{m_\rho^3}{f_\pi^2}\sqrt{\frac{2}{\pi}}\frac{3}{8I}.
\label{nucleon}
\ee
Taking the previously determined value 
$m_\rho=176\mbox{\,MeV}$ and requiring that (\ref{nucleon})
reproduces the nucleon mass $M_N=939\mbox{\,MeV}$ yields
$f_\pi=55.1\mbox{\,MeV},$ which is around $60\%$ of its
physical value. With these parameter values the classical and
quantum spin contributions to the nucleon mass are
$905\mbox{\,MeV}$ and $34\mbox{\,MeV}$ respectively, which is
an acceptable split. 

By equation (\ref{skyrmee}), 
this set of parameter values gives a
Skyrme parameter $e=3.81.$ It is interesting to note that these 
parameters are reasonably close to those
suggested in the usual Skyrme model by fitting to
the properties of the lithium-6 nucleus, which give
$f_\pi=37.6\mbox{\,MeV}$ and $e=3.26$ \cite{MW}.

\section{Conclusion}\news
Inspired by methods of holographic QCD, a sequence of extended Skyrme
models has been introduced that interpolate between the usual Skyrme
model and a BPS Skyrme model. This provides an explanation and extension
of the Atiyah-Manton construction of Skyrme fields from instanton holonomies,
as this construction produces exact solutions of the BPS Skyrme model.

The first extended Skyrme model is a nonlinear theory of pions 
coupled to the $\rho$ meson and this has been investigated in some detail. 
The results reveal that this model is significantly closer to a BPS theory 
than the usual Skyrme model, and it has been demonstrated that 
an extension of the Atiyah-Manton
construction provides an excellent approximation to both the pion
and $\rho$ meson fields. This is encouraging as the experimental
data on nuclear binding energies reveals that they are typically less than
$1\%,$ suggesting that a model close to a BPS theory is required.

There are several avenues for future research that follow from the work
presented in this paper. An obvious next step is to investigate
multi-Skyrmions in the extended theory, to confirm and evaluate
the reduced binding energies. This could be performed either using 
numerical full field simulations, similar to those applied in the
usual Skyrme model \cite{BS-full}, or using the instanton approximation.
A study of the extended Skyrme models that include additional vector
mesons would also be of interest: in particular
it would be useful to compute the change in binding energies as 
more vector mesons are introduced.

Many aspects of the Sakai-Sugimoto theory have been investigated 
since its introduction
and it would be interesting to attempt similar studies for the 
flat space analogue introduced here.
Examples of aspects to study include finite baryon density
and temperature \cite{HT,BLL,KSZ,RSRW,ASY,PSZ}, 
the determination of interaction coupling constants and form factors
\cite{HRYY,KZ,HRYYc,HSS}, together with an analysis of the nuclear force
\cite{HSSb}. 

The Skyrme models considered 
in this paper are all applicable to massless pions,
but it has been shown that in the usual Skyrme model there are significant
differences in the massive pion theory, and the differences are 
encouraging in respect
to comparisons with the properties of nuclei \cite{BS-pm,BMS,BMSW}.
A pion mass term could simply be included in the extended Skyrme models,
although the connection to a BPS Skyrme theory would then be lost. However,
it is still to be expected that binding energies would be reduced in 
comparison to the usual Skyrme model, as the resulting increase in energy 
applies to Skyrmions of all baryon numbers. The Atiyah-Manton 
construction does not produce Skyrme fields with asymptotic fields
appropriate to massive pions, but a modification of this 
construction has been introduced for massive pions \cite{AS}, 
based on a connection to hyperbolic Skyrmions. It would be interesting to
see if this modified construction can be understood in terms of the 
techniques introduced in the present paper, and perhaps this might lead
to an extended BPS Skyrme theory for massive pions.

Finally, during the preparation of this manuscript a preprint appeared
\cite{ASW} introducing novel Skyrmions in a different BPS Skyrme model. 
The BPS Skyrme model in question 
is of the type introduced some time ago \cite{ARTY},
involving only a pion mass term and a term of sixth-order in the derivatives of the
Skyrme field, and is therefore quite different from the one considered in the
present paper.
The novel Skyrmions presented in \cite{ASW}
are examples of compactons, that
is, they have compact support. Such Skyrmions are trivially BPS solitons,
since compactons placed far enough apart do not interact at all. 
However, there is an interesting mathematical structure, associated
with an infinite dimensional symmetry, that for example allows the single
Skyrmion to be obtained as a solution of a first order equation.

\newpage
\section*{Acknowledgements}
Many thanks to Nick Manton, Kasper Peeters and Marija Zamaklar 
for useful comments.

\noindent I acknowledge the STFC for support under the 
rolling grant ST/G000433/1.

\end{document}